\documentclass[a4paper]{article}%
\usepackage{amssymb,amsmath,amsfonts}
\usepackage{amsmath}
\usepackage{amsfonts}
\usepackage{amssymb}
\usepackage{graphicx}%
\providecommand{\U}[1]{\protect\rule{.1in}{.1in}}
\newtheorem{theorem}{Theorem}

\newtheorem{idea memo}[theorem]{Idea Memo}

\newtheorem{remark}[theorem]{Remark}

\begin{document}

\title{Perspective from Micro-Macro Duality\thanks{Talk at the International
Symposium, QBIC2008}\\--Towards non-perturbative \\renormalization scheme--}
\author{Izumi OJIMA\\RIMS, Kyoto University}
\date{}
\maketitle

\begin{abstract}
The problem of renormalization procedure is re-examined from the viewpoint of
Micro-Macro duality.

\end{abstract}

\section{Micro-Macro duality}

\textquotedblleft Micro-Macro duality\textquotedblright\ is one of the basic
features found between the invisible microscopic nature and its visible
macroscopic manifestations, which can be understood in parallel with the
Fourier duality between an abstract group and the concrete representations of
the former. This viewpoint has played crucial roles in our analysis of the
mutual relations between virtual dynamical levels and specific geometric ones
in various contexts (see, \cite{Oji05}). Using this general notion, we can
provide the heuristic idea of \textquotedblleft Quantum-Classical
Correspondence\textquotedblright\ with precise mathematical formulations in
which Micro and Macro are mutually and closely related with each other; the
latter, Macro, emerges from the former, Micro, through the processes of
condensation of infinitely many quanta and the essential features of the
former can be determined and re-constructed to certain extent from the data
structure at the levels of Macro, in close analogy with the above-mentioned
duality in the context of groups and representations. From this viewpoint of
Micro-Macro duality, we try here to sketch the essential ingredients for a
natural reformulation of the traditional theory of renormalization procedures
commonly adopted in the physical applications of quantum field theory (QFT for
short). For this purpose, the most relevant notions in what follows are the
group of scale transformations and the associated aspects of \textit{broken
symmetry} which is not unitarily implementable within a sector (defined by a
quasi-equivalence class of factor representations of the algebra of
observables) but which generates a family of (mutually disjoint) sectors along
an orbit of symmetry transformations.~

\section{Broken scale invariance: imaginary-time vs. real-time}

Here we briefly summarize some consequences of broken scale invariance in
relativistic QFT.

\vskip14pt 1) Imaginary-time version = \textquotedblleft temperature as order
parameter of broken scale invariance\textquotedblright:

\begin{theorem}
[IO04 \cite{Oji04}]In the standard setting up of algebraic QFT, the inverse
temperature $\beta:=(\beta^{\mu}\beta_{\mu})^{1/2}$ is a macroscopic
\textbf{order parameter }for parametrizing mutually \textit{disjoint} sectors
in the thermal situation arising from the \textbf{broken scale invariance}
under the renormalization-group transformations, where $\beta^{\mu}$ is an
inverse temperature 4-vector of a relativistic KMS state $\omega_{\beta^{\mu}%
}$ describing a thermal equilibrium in its rest frame.
\end{theorem}

This result is based on the notion of a scaling algebra due to Buchholz-Verch
\cite{BucVer} in combination with Takesaki's theorem \cite{Tak70} on the
disjointness of KMS states at different temperatures valid for a system with
physical observables constituting a von Neumann algebra of type III. \vskip14pt

2) What should be the corresponding \textquotedblleft
real-time\textquotedblright\ version to the above?: Renormalization Theory (at
$T=0K$).

\subsection{How to formulate broken scale invariance}

\begin{theorem}
[Takesaki'70 \cite{Tak70}]For a quantum C*-dynamical system with \textit{type
III} representations in its KMS states, any pair of KMS states for different
(inverse) temperatures $\beta_{1}\neq\beta_{2}$ are mutually \textit{disjoint}
$\omega_{\beta_{1}}\overset{\shortmid}{\circ}\omega_{\beta_{2}}$.
\end{theorem}

The claim of the first theorem due to myself is that the above disjointness
allows us to interpret the inverse temperature $\beta$ as an order parameter
of broken scale invariance. In the usual situation, this kind of symmetry
breakdown arises as a \textit{spontaneous breakdown} of a symmetry described
by a group acting on the algebra of physical quantities by automorphisms. In
contrast, the present case of broken scale invariance usually involves
\textit{explicit breaking terms }such as \textit{mass}, which seem to
\textit{prevent scale transformations from being treated as automorphisms.
}However, the results on \textit{scaling algebra} in algebraic QFT due to
\cite{BucVer, Oji04} shows that the above negative anticipation can be avoided.

Their results can be summarized as follows. Let the following requirements be
imposed on all the possible renormalization-group transformations $R_{\lambda
}$:

(i) $R_{\lambda}$ should map the given net $\mathcal{O}\rightarrow
\mathcal{A}(\mathcal{O})$ of local observables at spacetime scale $1$ onto the
corresponding net $\mathcal{O}\rightarrow\mathcal{A}_{\lambda}(\mathcal{O}%
)\doteq\mathcal{A}({\lambda}\mathcal{O})$ at scale $\lambda$, i.e.,
\[
R_{\lambda}:\,\mathcal{A}(\mathcal{O})\rightarrow\mathcal{A}_{\lambda
}(\mathcal{O})
\]
for every region $\mathcal{O}\subset\mathbb{R}^{4}$, through which the light
velocity $c$ is kept unchanged: $(\lambda x)^{i}/(\lambda x)^{0}=x^{i}/x^{0}.$

(ii) In the Fourier-transformed picture, the subspace $\widetilde{\mathcal{A}%
}(\widetilde{\mathcal{O}})$ of all (quasi-local) observables carrying
energy-momentum in the set $\widetilde{\mathcal{O}}\subset\mathbb{R}^{4}$ is
transformed as
\[
R_{\lambda}:\,\widetilde{\mathcal{A}}(\widetilde{\mathcal{O}})\rightarrow
\widetilde{\mathcal{A}}_{\lambda}(\widetilde{\mathcal{O}})
\]
for $\forall\widetilde{\mathcal{O}}$, where $\widetilde{\mathcal{A}}_{\lambda
}(\widetilde{\mathcal{O}}):=\widetilde{\mathcal{A}}(\lambda^{-1}%
\widetilde{\mathcal{O}})$, through which the Planck constant $\hbar$ is
unchanged: $(\lambda^{-1}p)_{\mu}(\lambda x)^{\mu}/\hbar=p_{\mu}x^{\mu}/\hbar$.

(iii) For scale invariant theories $R_{\lambda}$ may not be isomorphisms but
are maps continuous and bounded uniformly in $\lambda$.

Then, \textit{scaling net }$\mathcal{O}\rightarrow\widehat{\mathcal{A}%
}(\mathcal{O})$ corresponding to the original local net $\mathcal{O}%
\rightarrow\mathcal{A}(\mathcal{O})$ of observables is defined as the local
net consisting of scale-changed observables under the action of all the
possible choice of $R_{\lambda}$ satisfying (i)-(iii). Mathematically the
algebra $\widehat{\mathcal{A}}(\mathcal{O})$ can be understood as the algebra
$\Gamma(\mathbb{R}^{+}\times\mathcal{A}(\mathcal{O}))$ of sections
$\mathbb{R}^{+}\ni\lambda\longmapsto\hat{A}(\lambda)\in\mathcal{A}_{\lambda
}(\mathcal{O})$ of algebra bundle $\amalg_{\lambda\in\mathbb{R}^{+}%
}\mathcal{A}_{\lambda}(\mathcal{O})\twoheadrightarrow\mathbb{R}^{+}$ over the
multiplicative group $\mathbb{R}^{+}$ of scale changes ($=$ $\mathcal{A}%
(\mathcal{O})\rtimes_{\alpha}\widehat{\mathbb{R}^{+}}$: augmented algebra).
Then, the \textit{scaling algebra} $\hat{\mathcal{A}}$ is defined by the
C*-inductive limit of all local algebras $\hat{\mathcal{A}}(\mathcal{O}%
)$.~Algebraic structures making $\hat{\mathcal{A}}(\mathcal{O})$ a unital
C*-algebra are defined in a \textit{pointwise manner}, for instance, by
$(\hat{A}\cdot\hat{B})(\lambda):=\hat{A}(\lambda)\hat{B}(\lambda)$, $(\hat
{A}^{\ast})(\lambda):=\hat{A}(\lambda)^{\ast}$, etc., and $||\,\hat
{A}\,||:=\sup_{\lambda\in\mathbb{R}^{+}}\,||\,\hat{A}(\lambda)\,||$.

From the scaled actions $\mathcal{A}_{\lambda}\underset{\alpha^{(\lambda)}%
}{\curvearrowleft}\mathcal{P}_{+}^{\uparrow}$ of Poincar\'{e} group on
$\mathcal{A}_{\lambda}$ with $\alpha_{x,\Lambda}^{(\lambda)}=\alpha_{\lambda
x,\Lambda}$, an action of $\mathcal{P}_{+}^{\uparrow}$ is induced on
$\widehat{\mathcal{A}}$ by
\[
(\hat{\alpha}_{x,\Lambda}(\hat{A}))(\lambda):=\alpha_{\lambda x,\Lambda}%
(\hat{A}(\lambda))\,.
\]
Then the conditions (ii), (iii) are expressed simply as the continuity of
Poincar\'{e}-group action: $||\,\hat{\alpha}_{x,\Lambda}(\hat{A})-\hat
{A}\,||\underset{(x,\Lambda)\rightarrow(0,1)}{\rightarrow}0$. Then, the
scaling net\textit{\ }$\mathcal{O}\rightarrow\widehat{\mathcal{A}}%
(\mathcal{O})$ is shown to satisfy all the properties to characterize a
relativistic local net of observables if the original one $\mathcal{O}%
\rightarrow\mathcal{A}(\mathcal{O})$ does.

Now scale transformations can be defined by an automorphic action $\hat
{\sigma}_{\mathbb{R}{^{+}}}$ of the $\mathbb{R}^{+}$ on the scaling algebra
$\hat{\mathcal{A}}$, given for $\forall\mu\in\mathbb{R}^{+}$ by
\[
(\hat{\sigma}_{\mu}(\hat{A}))(\lambda):=\hat{A}(\mu\lambda),\quad\lambda>0,
\]
satisfying
\begin{align*}
\hat{\sigma}_{\mu}(\hat{\mathcal{A}}(\mathcal{O}))  &  =\hat{\mathcal{A}}%
(\mu\mathcal{O})\,,\quad\mathcal{O}\subset\mathbb{R}^{4},\\
\hat{\sigma}_{\mu}\circ\hat{\alpha}_{x,\Lambda}  &  =\hat{\alpha}_{\mu
x,\Lambda}\circ\hat{\sigma}_{\mu}\,,\quad(x,\Lambda)\in\mathcal{P}%
_{+}^{\uparrow}\,.
\end{align*}

\begin{remark}
Scaling transformations $\hat{\sigma}_{\mathbb{R}{^{+}}}$ play the role of
renormalization group transformations to relate observables at different scales.
\end{remark}

\begin{remark}
Since a broken symmetry can always be restored by taking all breaking
parameters as variables undergoing the broken symmetry transformations, there
is no miracle in the results due to Buchholz and Verch through their
complicated analysis: it can naturally be accommodated as a special case into
the general definition of a augmented algebra \cite{Unif03} $\mathcal{\hat{F}%
}:=\Gamma(G\times_{H}\mathcal{F})$ with the choice of $H:=\mathcal{P}%
_{+}^{\uparrow}$, $G=H\rtimes\mathbb{R}^{+}$ (semidirect product) and together
with slight modifications due to spacetime dependence $\mathcal{F}%
\Longrightarrow\mathcal{A}(\mathcal{O})$ (which is affected by the action of
$\mathbb{R}^{+}$) (and the intervention of the centre due to SSB:
$SO(3)\backslash L_{+}^{\uparrow}\cong\mathbb{R}^{3}$ at $T\neq0^{\circ}K$)
\cite{Oji04}. \newline\ \ Scaled actions $\alpha_{x,\Lambda}^{(\lambda
)}=\alpha_{\lambda x,\Lambda}$ of Poincar\'{e} group on $\mathcal{A}_{\lambda
}$ can also be naturally understood as the conjugacy change of the stability
group $H\rightarrow gHg^{-1}$ from the point $He$ to $Hg^{-1}$ on the base
space $H\backslash G=\mathbb{R}^{+}$: $s_{\mu}(x,\Lambda)s_{\mu}^{-1}=(\mu
x,\Lambda)$.
\end{remark}

\subsection{Scale changes on states}

Corresponding to each probability measure $\mu$ on the centre $\mathfrak{Z}%
(\widehat{\mathcal{A}})=\mathfrak{Z}(\widehat{\mathcal{A}}(\mathcal{O}%
))=C(\mathbb{R}^{+})$ due to the broken scale invariance, we have a
conditional expectation $\hat{\mu}$ from $\widehat{\mathcal{A}}$ to
$\mathcal{A}$:
\begin{equation}
\hat{\mu}:\widehat{\mathcal{A}}\ni\hat{A}\longmapsto\int_{\mathbb{R}^{+}}%
d\mu(\lambda)\hat{A}(\lambda)\in\mathcal{A}\mathfrak{.}%
\end{equation}
Instead of $d\mu(\lambda)$, it is also possible to take the Haar measure
$d\lambda/\lambda$ of $\mathbb{R}^{+}$. As it is a positive unbounded measure
but not a probability one with the total mass one, however, the corresponding
map to $\hat{\mu}$ becomes an \textit{operator-valued weight }whose images are
not guaranteed to be finite. Any state $\omega\in E_{\mathcal{A}}$ can be
lifted onto $\widehat{\mathcal{A}}$ through $\hat{\mu}$ by
\begin{equation}
E_{\mathcal{A}}\ni\omega\longmapsto\hat{\mu}^{\ast}(\omega)=\omega\circ
\hat{\mu}=\omega\otimes\mu\in E_{\widehat{\mathcal{A}}},
\end{equation}
where we have used $\widehat{\mathcal{A}}\subset C(\mathbb{R}^{+}%
,\mathcal{A})\cong\mathcal{A}\otimes C(\mathbb{R}^{+})$.

In \cite{BucVer}{\ the case $\mu=\delta_{\lambda=1}$(: Dirac measure at the
identity of $\mathbb{R}^{+}$) is called a \textit{canonical lift }$\hat
{\omega}:=\omega\circ\hat{\delta}_{1}$. The scale transformed state defined
by
\begin{equation}
\hat{\omega}_{\lambda}:=\hat{\omega}\circ\hat{\sigma}_{\lambda}=\omega
\circ\hat{\delta}_{\lambda}%
\end{equation}
describes the situation at scale $\lambda$ due to the renormalization-group
transformation of scale change $\lambda$. }

Conversely, starting from a state $\hat{\omega}$ of $\widehat{\mathcal{A}}$,
we can obtain its central decomposition as follows: first, we call two natural
embedding maps $\iota:\mathcal{A}\mathfrak{\hookrightarrow}\widehat
{\mathcal{A}}$ [$\left[  \iota(A)\right]  (\lambda)\equiv A$] and
$\kappa:C(\mathbb{R}^{+})\simeq\mathfrak{Z}(\widehat{\mathcal{A}%
})\hookrightarrow\widehat{\mathcal{A}}$. Pulling back $\hat{\omega}$ by
$\kappa^{\ast}:E_{\widehat{\mathcal{A}}}\rightarrow E_{C(\mathbb{R}^{+})}$, we
can define a probability measure $\rho_{\hat{\omega}}:=\kappa^{\ast}%
(\hat{\omega})=\hat{\omega}\circ\kappa=$\ $\hat{\omega}\upharpoonright
_{C(\mathbb{R}^{+})}$ on $\mathbb{R}^{+}$, namely, $\hat{\omega}%
\upharpoonright_{C(\mathbb{R}^{+})}(f)=\int_{\mathbb{R}^{+}}d\rho_{\hat
{\omega}}(\lambda)f(\lambda)$ for $\forall f\in$ $C(\mathbb{R}^{+})$.

For any positive operator $\hat{A}=\int ad\hat{E}_{\hat{A}}(a)\in
\widehat{\mathcal{A}}$, we can consider the central supports $c(\hat{E}%
_{\hat{A}}(\Delta))\in Proj(\mathfrak{Z}_{\hat{\pi}_{\hat{\omega}}}%
(\widehat{\mathcal{A}}))$ of $\hat{E}_{\hat{A}}(\Delta)\in Proj(\hat{\pi
}_{\hat{\omega}}(\widehat{\mathcal{A}})^{\prime\prime})$ with a Borel set
$\Delta$ in $Sp(\hat{A})\subset\lbrack0,+\infty)$ satisfying $c(\hat{E}%
_{\hat{A}}(\Delta))\hat{E}_{\hat{A}}(\Delta)=\hat{E}_{\hat{A}}(\Delta)$. From
this we see that $\rho_{\hat{\omega}}^{\prime\prime}(c(\hat{E}_{\hat{A}%
}(\Delta)))=0$ implies $\hat{\omega}^{\prime\prime}(\hat{E}_{\hat{A}}%
(\Delta))=0$, where $\hat{\omega}^{\prime\prime}$ and $\rho_{\hat{\omega}%
}^{\prime\prime}$ are the extensions of $\hat{\omega}$ and $\rho_{\hat{\omega
}}$ to $\hat{\pi}_{\hat{\omega}}(\widehat{\mathcal{A}})^{\prime\prime}$ and
$L^{\infty}(\mathbb{R}^{+},d\rho_{\hat{\omega}})$, respectively. Thus, we can
define the Radon-Nikodym derivative $\omega_{\lambda}:=\frac{d\hat{\omega}%
}{d\rho_{\hat{\omega}}}(\lambda)$ of $\hat{\omega}$ w.r.t.~$\rho_{\hat{\omega
}}$ as a state on $\hat{\pi}_{\hat{\omega}}(\widehat{\mathcal{A}}%
)^{\prime\prime}$ so that%

\begin{equation}
\hat{\omega}(\hat{A})=\int d\rho_{\hat{\omega}}(\lambda)\omega_{\lambda}%
(\hat{A}(\lambda))=\int d\rho_{\hat{\omega}}(\lambda)\omega_{\lambda}%
(\hat{\delta}_{\lambda}(\hat{A}))=\int d\rho_{\hat{\omega}}(\lambda)\left[
\omega_{\lambda}\otimes\hat{\delta}_{\lambda}\right]  (\hat{A}).
\end{equation}
Then, the pull-back $\iota^{\ast}(\hat{\omega})=\hat{\omega}\circ\iota\in
E_{\mathcal{A}}$ of $\hat{\omega}\in E_{\widehat{\mathcal{A}}}$ by
$\iota^{\ast}:E_{\widehat{\mathcal{A}}}\rightarrow E_{\mathcal{A}}$ is given
by
\begin{equation}
\iota^{\ast}(\hat{\omega})=\int d\rho_{\hat{\omega}}(\lambda)\omega_{\lambda},
\end{equation}
owing to the relation%

\begin{equation}
\iota^{\ast}(\hat{\omega})(A)=\hat{\omega}(\iota(A))=\int d\rho_{\hat{\omega}%
}(\lambda)\omega_{\lambda}(A)=\left[  \int d\rho_{\hat{\omega}}(\lambda
)\omega_{\lambda}\right]  (A).
\end{equation}
Using this relation to the scaled canonical lift, $\hat{\omega}_{\lambda
}:=\hat{\omega}\circ\hat{\sigma}_{\lambda}=(\omega\circ\hat{\delta}_{1}%
)\circ\hat{\sigma}_{\lambda}=\omega\circ\hat{\delta}_{\lambda}$, of a state
$\omega\in E_{\mathcal{A}}$, we can easily see $\iota^{\ast}(\omega\circ
\hat{\delta}_{\lambda})=\iota^{\ast}(\hat{\omega}_{\lambda})=\omega_{\lambda
}[=\frac{d\hat{\omega}_{\lambda}}{d\delta_{\lambda}}(\lambda)]=\phi_{\lambda
}(\omega)$, where $\phi_{\lambda}$ is the isomorphism introduced in
\cite{BucVer} between $\omega$ and the canonical lift $\hat{\omega}_{\lambda
}\in E_{\widehat{\mathcal{A}}}$ projected onto $\widehat{\mathcal{A}%
}/\mathrm{ker}(\hat{\pi}_{\hat{\omega}}\circ\hat{\sigma}_{\lambda})$.

Thus we can lift any state $\omega\in E_{\mathcal{A}}$ canonically from
$\mathcal{A}$ to $\hat{\omega}\in E_{\widehat{\mathcal{A}}}$, and, after the
scale shift $\hat{\sigma}_{\lambda}$ on $\widehat{\mathcal{A}}$, return
$\hat{\omega}\circ\hat{\sigma}_{\lambda}$ back onto $\mathcal{A}$:
$\phi_{\lambda}(\omega)=\omega_{\lambda}=\iota^{\ast}(\omega\circ\hat{\delta
}_{\lambda})$, as result of which we obtain the scaled-shifted state
$\omega_{\lambda}\in E_{\mathcal{A}}$ from $\omega\in E_{\mathcal{A}}$
\textit{in spite of the absence of scale invariance on} $\mathcal{A}$.

Now applying this procedure to $\omega=\omega_{\beta}$ (: any state belonging
to the family of relativistic KMS states with the same $(\beta^{2})^{1/2}$),
we have a genuine KMS state by going to their rest frames. Then we have
$\hat{\omega}_{\lambda}=(\widehat{\omega_{\beta}})_{\lambda}=\omega_{\beta
}\circ\widehat{\delta_{\lambda}}$ which is shown to be a KMS state at
$\beta/\lambda$:%
\begin{align}
(\omega_{\beta}\circ\widehat{\delta_{\lambda}})(\hat{A}\hat{\alpha}_{t}%
(\hat{B}))  &  =\omega_{\beta}(\hat{A}(\lambda)\alpha_{\lambda t}(\hat
{B}(\lambda)))\nonumber\\
=\omega_{\beta}(\alpha_{\lambda t-i\beta}(\hat{B}(\lambda))\hat{A}(\lambda))
&  =\omega_{\beta}(\alpha_{\lambda(t-i\beta/\lambda)}(\hat{B}(\lambda))\hat
{A}(\lambda))\nonumber\\
&  =(\omega_{\beta}\circ\widehat{\delta_{\lambda}})(\hat{\alpha}%
_{t-i\beta/\lambda}(\hat{B})\hat{A}),
\end{align}
and hence, $(\widehat{\omega_{\beta}})_{\lambda}\in\hat{K}_{\beta/\lambda}$,
$\phi_{\lambda}(\omega_{\beta})\in K_{\beta/\lambda}$.

As already remarked, the above discussion is seen to apply equally to the
spontaneous as well as \textit{explicitly broken} scale invariance with
\textit{explicit breaking} parameters such as mass terms. The actions of scale
transformations on such variables as $x^{\mu}$, $\beta^{\mu}$ and also
conserved charges are just straightforward, which is justified by such facts
that the first and the second ones are of \textit{kinematical} nature and that
the second and the third ones exhibit themselves in the \textit{state labels}
for specifying the relevant \textit{sectors} in the context of the
superselection structures \cite{Oji03, Unif03}. This gives an alternative
verification to the so-called \textit{non-renormalization theorem of conserved
charges}. In sharp contrast, other such variables as coupling constants (to be
read off from the data of correlation functions or Green's functions) are
affected by the \textit{scaled dynamics,} and hence, may show non-trivial
scaling behaviours with deviations from the canonical (or kinematical)
dimensions, in such forms as the running couplings or anomalous dimensions.
Thus, the transformations $\hat{\sigma}_{\lambda}$ (as \textquotedblleft
exact\textquotedblright\ symmetry on the augmented algebra $\widehat
{\mathcal{A}}$) are understood to play the roles of the renormalization-group
transformations (as broken symmetry on the original algebra $\mathcal{A}$). As
a result, we see that \textit{classical macroscopic observable} $\beta$
naturally emerging from a microscopic quantum system is verified to be an
\textit{order parameter }of\textit{\ broken scale invariance} involved in the
renormalization group.

In the present context of the scale transformations in real version, we can
use these scale changes of states to compare different theories renormalized
by renormalization conditions imposed at different scale points.

\section{Nuclearity Condition \& Renormalizability}

For the purpose of controlling the phase space properties in algebraic QFT,
the \textit{nuclearity condition} is formulated as follows: the map
$\Phi_{\mathcal{O},E}:\mathcal{A}(\mathcal{O})_{1}\ni A\longmapsto
P_{E}A\Omega\in\mathfrak{H}$ with $P_{E}$ the spectral projection on state
vectors having energy below $E$ is required to be \textit{nuclear}, admitting
such a decomposition as%
\begin{align*}
\Phi_{\mathcal{O},E}(A)  &  =\sum_{i=1}^{\infty}\varphi_{i}(A)\xi_{i}\text{
\ \ \ for }\forall A\in\mathcal{A}(\mathcal{O})_{1}\text{ }\\
\text{with }\varphi_{i}  &  \in\mathcal{A}(\mathcal{O})^{\ast}\text{ and }%
\xi_{i}\in\mathfrak{H}\text{ s.t. }\sum_{i=1}^{\infty}\left\vert \left\vert
\varphi_{i}\right\vert \right\vert \left\vert \left\vert \xi_{i}\right\vert
\right\vert <\infty,
\end{align*}
on the unit ball $\mathcal{A}(\mathcal{O})_{1}:=\{A\in\mathcal{A}%
(\mathcal{O});$ $\left\vert \left\vert A\right\vert \right\vert \leq1\}$ of
any local subalgebra $\mathcal{A}(\mathcal{O})$ of observables. This condition
excludes such \textquotedblleft unphysical\textquotedblright\ fields as
\textit{generalized free fields} without discrete mass spectrum admitting no
\textit{particle picture} to be detected in scattering experiments. The
nuclearity condition and the\ assumption of the approximate scale invariance
are known \cite{Fre85} to imply that the local subalgebra $\mathcal{A}%
(\mathcal{O})$ is a factor von Neumann algebra of \textit{type III} with
\textit{no minimal projections}, i.e., any projection operator $E\in
\mathcal{A}(\mathcal{O})\diagdown\{0\}$ is equivalent to the identity operator
$I=id_{\mathfrak{H}}$: $\exists v\in\mathcal{A}(\mathcal{O})$ s.t. $v^{\ast
}v=I,vv^{\ast}=E$.

\subsection{Point-like fields as idealized local observables}

On the basis of the nuclearity condition \cite{NuclCond} and the energy-bound
, the notion of point-like field operators \cite{FrHe, HaOj} has been
established satisfying the operator-product expansion (\textit{OPE}) in a
\textit{non-perturbative} way in algebraic QFT by Bostelmann \cite{Bo}. The
\textit{energy bound} means the requirement that observed values of quantum
fields $\hat{\phi}(f)$ can become large only with large energy: for any $l>0$,
there is a sufficiently large $m>0$ that the inequality
\[
||(1+H)^{-m}\,\hat{\phi}(f)\,(1+H)^{-m}||\leq c\,\int\!dx\,|(1-\Delta
)^{-l}f(x)|,
\]
holds with a (positive) Hamiltonian $H$, operator norm $||\cdot||$ in the
vacuum sector $\mathcal{H}$ and $\Delta$: Laplacian on $\mathbb{R}^{4}$. When
this holds, there exist a sequence of test functions tending to $f_{i}%
\underset{i\mathbb{\rightarrow}\infty}{\rightarrow}\delta_{x}$: Dirac measure
at $x$ and a sufficiently large integer $m>0$ such that
\[
\lim_{i\rightarrow\infty}\,(1+H)^{-m}\,\hat{\phi}(f_{i})\,(1+H)^{-m}%
=:(1+H)^{-m}\,\hat{\phi}(x)\,(1+H)^{-m}.
\]
Then a \textit{field} $\hat{\phi}(x)$ \textit{at a point} $x$ is well-defined
as a linear form on such states $\omega$ in the vacuum sector that
$\omega((1+H)^{2m})<\infty$. Hermitian elements in the sets $\mathcal{Q}%
_{\,m,x}:=\{\hat{\phi}(x);$ $||(1+H)^{-m}\hat{\phi}(x)(1+H)^{-m}||<\infty\}$
of point-like fields are idealized observables at spacetime points $x$
meaningful for such states $\omega$ that $\omega((1+H)^{2m})<\infty$. The set
$\mathcal{Q}_{\,m,x}$ of such point-like fields are, in general,
finite-dimensional linear spaces satisfying $\mathcal{Q}_{\,m,x}%
\subset\mathcal{Q}_{\,m^{\prime},x}$ for $m\leq m^{\prime}$ and are invariant
under the stability group of $x$ in $\mathcal{P}_{+}^{\uparrow}$. The
meaningless notion of \textit{product of fields at a point\ }$x$ is replaced
in $\mathcal{Q}_{\,m,x}$ by \textit{normal products} defined by the following
OPE: for instance, ill-defined square $\hat{\phi}(x)^{2}$ is replaced by the
subspaces $\mathcal{N}(\hat{\phi}^{2})_{\,q,x}\subset\mathcal{Q}_{\,n,x}$
generated by normal products $\hat{\Phi}_{j}(x)$, $j=1,\cdots,J(q)$, appearing
in OPE of $\hat{\phi}(x+\frac{\xi}{2})\hat{\phi}(x-\frac{\xi}{2})$:
\[
||(1+H)^{-n}\left[  \hat{\phi}(x+\frac{\xi}{2})\hat{\phi}(x-\frac{\xi}%
{2})-\sum_{j=1}^{J(q)}c_{j}(\xi)\,\hat{\Phi}_{j}(x)\right]  (1+H)^{-n}||\leq
c\,|\xi|^{q},
\]
which is satisfied for any $\hat{\phi}\in\mathcal{Q}_{\,m,x}$ for spacelike
$\xi(\in\mathbb{R}^{4})\rightarrow0$ with arbitrary $q>0$, by choosing a
finite number of fields $\hat{\Phi}_{j}(x)\in\mathcal{Q}_{\,n,x}$ and
sufficiently large $n$, and some analytic functions $\xi\mapsto c_{j}(\xi)$,
$j=1,\cdots,J(q)$. Using this definition, the spaces $\mathcal{N}(\hat{\phi
}^{\,p})_{\,q,x}(\subset\mathcal{Q}_{\,n,x})$ of normal products of higher
powers $p$ can similarly be defined. While the linear spaces $\mathcal{Q}%
_{\,m,x}$ of pointlike fields lack the multiplication structure, the validity
of OPE allows us to provide them with a structure generalizing
a\textit{\ product system of\ Hilbert modules }$\mathcal{Q}_{\,n,x}$ . It is
also possible for the partial derivatives $\partial_{\xi}$ of spacetime
coordinates $\xi$ to act on these spaces through the \textquotedblleft%
\textit{balanced\ derivatives\textquotedblright} $\partial_{\xi}\,\hat{\phi
}(x+\frac{\xi}{2})\hat{\phi}(x-\frac{\xi}{2})$ \cite{BOR01} which are
contained in $\mathcal{N}(\hat{\phi}^{2})_{\,q,x}$ (for large $q$) as shown by
the relation,%
\[
||(1+H)^{-n}[\partial_{\xi}\,\hat{\phi}(x+\frac{\xi}{2})\hat{\phi}(x-\frac
{\xi}{2})-\sum_{j=1}^{J(q)}\partial_{\xi}\,c_{j}(\xi)\ \hat{\Phi}%
_{j}(x)](1+H)^{-n}||\leq c\,|\xi|^{r},
\]
valid for $\forall r>0$ $\exists q$ and $\exists n$ sufficiently large.

\subsection{Comparison between OPE \& Wigner-Eckhart theorem}

To take advantage of the above OPE structure, we compare it with the basic
feature of the Wigner-Eckhart theorem for an irreducible family of tensor
operators $\{F_{m_{1}}^{(\gamma_{1})};m_{1}=-\gamma_{1},-\gamma_{1}%
+1,\cdots,\gamma_{1}-1,\gamma_{1}\}$ under the action of a (compact) group $G$
(such as $SU(2)$, typically):
\[
\langle\gamma m|F_{m_{1}}^{(\gamma_{1})}|\gamma_{2}m_{2}\rangle=\langle
\gamma||F^{(\gamma_{1})}||\gamma_{2}\rangle\langle\gamma m|(\gamma_{1}%
m_{1}),(\gamma_{2}m_{2})\rangle,
\]
where $\langle\gamma m|(\gamma_{1}m_{1}),(\gamma_{2}m_{2})\rangle$ are the
Clebsch-Gordan coefficients describing a branching rule from the Kronecker
tensor product $[\gamma_{1}\hat{\otimes}\gamma_{2}](g)=\gamma_{1}%
(g)\otimes\gamma_{2}(g)$ of representations $(\gamma_{i},V_{\gamma_{i}})$
($i=1,2$) into irreducible ones $\{(\gamma,V_{\gamma})\}\in Rep(G)$
($|\gamma,m\rangle\in V_{\gamma}$) of $G$.~Note that the Kronecker tensor
product $\gamma_{1}\hat{\otimes}\gamma_{2}$ of $G$ is the restriction of the
tensor product representation $(\gamma\boxtimes\gamma_{2})(g_{1},g_{2}%
)=\gamma(g_{1})\otimes\gamma_{2}(g_{2})$ of $G\times G$ onto a subgroup $G$
embedded via the diagonal map $\delta_{G}:G\ni g\longmapsto\delta
_{G}(g)=(g,g)\in G\times G$:
\[
\lbrack\gamma_{1}\hat{\otimes}\gamma_{2}](g)=[(\gamma_{1}\boxtimes\gamma
_{2})\circ\delta_{G}](g)=\gamma_{1}(g)\otimes\gamma_{2}(g).
\]
According to this formula, the matrix elements of the tensor operator
$\{F_{m_{1}}^{(\gamma_{1})};m_{1}=-\gamma_{1},-\gamma_{1}+1,\cdots,\gamma
_{1}-1,\gamma_{1}\}$ are decomposed into two factors, $G$-\textit{invariant
dynamical} one $\langle\gamma||F^{(\gamma_{1})}||\gamma_{2}\rangle$ \& purely
\textit{kinematical} one $\langle\gamma m|(\gamma_{1}m_{1}),(\gamma_{2}%
m_{2})\rangle$ determined completely by $G$-transformation property of
$F^{(\gamma_{1})}$.

In the case of OPE,
\[
\varphi_{1}(x+\frac{\xi}{2})\varphi_{2}(x-\frac{\xi}{2})\underset
{\xi\rightarrow0}{\thicksim}\sum_{i}N(\varphi_{1}\varphi_{2})_{i}(x)C_{i}%
(\xi)+\cdots,
\]
the dual map $\delta^{\ast}$ of $\delta$ given by $\delta^{\ast}(\varphi
_{1}\boxtimes\varphi_{2})(x)=(\varphi_{1}\boxtimes\varphi_{2})(\delta
(x))=\varphi_{1}(x)\otimes\varphi_{2}(x)$ is \textit{ill-defined} for
operator-valued distributions $\varphi_{i}$. In this context, therefore, the
diagonal map $\delta(x)=(x,x)$ should be understood in the limit:
$(x+\frac{\xi}{2},x-\frac{\xi}{2})\underset{\xi\rightarrow0}{\thicksim}%
\delta(x)=(x,x)$ after \textquotedblleft removing\textquotedblright\ such
divergent terms as $C_{i}(\xi)$. Except for this difference, the essence of
OPE formula is just in parallel with the above Wigner-Eckhart case:
\textit{factorization} of the product $\varphi_{1}(x+\frac{\xi}{2})\varphi
_{2}(x-\frac{\xi}{2})$ into two components, \textit{dynamical}
\textit{non-singular} factors $N(\varphi_{1}\varphi_{2})_{i}(x)$ depending
only on the \textquotedblleft centre of mass\textquotedblright\ $[(x+\frac
{\xi}{2})+(x-\frac{\xi}{2})]/2=x$ and \textit{c-number kinematical singular
functions} $C_{i}(\xi)$ of the relative coordinates $(x+\frac{\xi}%
{2})-(x-\frac{\xi}{2})=\xi$. Note that the \textit{singularity of the product
}$\varphi_{1}(x+\frac{\xi}{2})\varphi_{2}(x-\frac{\xi}{2})$\textit{\ in the
limit of }$\xi\rightarrow0 $ is isolated into these\textit{\ }kinematical
c-number factors $C_{i}(\xi)=N_{i}(\lambda)C_{i}^{reg}(\xi)$, where
$\lambda:=$ $\left\vert \xi\right\vert ^{-1}$ represents the \textit{cutoff
momentum} to regularize the \textit{UV divergences} in a
\textit{non-perturbative} way and $N_{i}(\lambda)$ can be taken as
\textit{counter terms} to define \textit{renormalized field operators}
(formally) by
\[
\varphi_{ren}(x):=\Pi_{i}N_{i}(\lambda)^{-1/2}\varphi(x).
\]
It may be instructive to find the analogy of the present structure with the
time-localization scale $\Delta t$ of Hida derivatives $a_{t},a_{t}^{\ast}$ in
the White-Noise Analysis \cite{2}.

Since the limit $\underset{\xi\rightarrow0}{\thicksim}$ means
\[
||(1+H)^{-n}\left[  \hat{\phi}(x+\frac{\xi}{2})\hat{\phi}(x-\frac{\xi}%
{2})-\sum_{j=1}^{J(q)}c_{j}(\xi)\,\hat{\Phi}_{j}(x)\right]  (1+H)^{-n}||\leq
c\,|\xi|^{q},
\]
the convergence $\hat{\phi}(x+\frac{\xi}{2})\hat{\phi}(x-\frac{\xi}%
{2})\rightarrow\sum_{j=1}^{J(q)}c_{j}(\xi)\,\hat{\Phi}_{j}(x)$ is
\textit{state-dependent} so that
\[
\omega(\left[  \hat{\phi}(x+\frac{\xi}{2})\hat{\phi}(x-\frac{\xi}{2}%
)-\sum_{j=1}^{J(q)}c_{j}(\xi)\,\hat{\Phi}_{j}(x)\right]  )\underset
{\xi\rightarrow0}{\rightarrow}0
\]
holds only for those states $\omega$ which satisfy
\[
\omega((1+H)^{2n})<constant.
\]
Thus, states $\omega$ for which OPE is valid \textit{cannot be localized}, and
hence, to such an extent, the spacetime point $x$ in $\varphi(x)$ is actually extended!

In the above situations the common essence can be found in the relevance of
some selective filters depending on the choice of states $\omega$. We note
here that the (approximate)\ \textit{diagonal maps} $\delta(x)=(x,x)$,
\begin{align*}
(x+\frac{\xi}{2},x-\frac{\xi}{2})\underset{\xi\rightarrow0}{\thicksim}(x,x)
&  =\delta(x);\\
\lbrack(\gamma_{1}\boxtimes\gamma_{2})\circ\delta](g)  &  =[\gamma_{1}%
\hat{\otimes}\gamma_{2}](g)=\gamma_{1}(g)\otimes\gamma_{2}(g),
\end{align*}
play essential roles in the definition of \textit{Hopf algebra} structures
with the harmonic-analytic dualities controlled by Kac-Takesaki operator (of
the so-called duality transformations), which should play crucial roles in
extending the above relations for two-point functions to arbitrary $n$-point functions.

By the above condition $\omega((1+H)^{2n})<constant$, the selective filter on
the initial state $\omega$ is related with the nuclearity condition
$\Phi_{\mathcal{O},E}(A)=P_{E}A\Omega=\sum_{i=1}^{\infty}\varphi_{i}(A)\xi
_{i}$ whose energy scale $E$ can be related to the above \textit{cutoff}
$\lambda$. In spite of the sharp contrast between \textit{almost
finite-dimensitonality} as nuclearity and $\infty$\textit{-dimensionality}
inherent to \textit{type III}, both properties are closely related with the
\textit{nuclearity condition} and are crucial for \textit{renormalizability}
and for shifts of the \textit{renormalization points} by
\textit{renormalization-group} transformations:

1) \textit{renormalizability} = finiteness of the number of graph types of
divergent \textquotedblleft1-particle irreducible (1PI)\textquotedblright%
\ diagrams is expected to follow from the very nuclearity condition ($=$
\textit{intra-sectorial} structure);

2) the \textit{absence of minimal projection in type III} von Neumann factors
(due to approximate scale invariance) allows the \textit{shifts of
renormalization points} by scale transformations =
\textit{renormalization-group transformations}. This gives the
\textit{inter-sectorial} relations among \textit{\textquotedblleft sectors
parametrized by renormalization conditions\textquotedblright\ at different
renormalization points} (on the centre $\mathfrak{Z}(\widehat{\mathcal{A}%
})=\mathfrak{Z}(\widehat{\mathcal{A}}(\mathcal{O}))=C(\mathbb{R}^{+})$ of the
scaling algebra).

In this sense, the \textit{nuclearity condition }can be regarded as the
\textit{mathematical version of the renormalizability condition} and
\textit{broken scale invariance }inherent to local subalgebras $\mathcal{A}%
(\mathcal{O})$ of\textit{\ type III} with no minimal projection requires the
\textit{renormalization condition to be specified at some renormalization
point} which can, however, be \textit{chosen arbitrarily}.

Here we present some new perspectives for understanding the conceptual and
mathematical meaning of \textit{renormalization scheme} in relation with such
key notions as the nuclearity condition, broken scale invariance and the type
III nature of local subalgebras of quantum fields in close relation with
algebraic QFT. What remains to be clarified is the following:

\begin{enumerate}
\item \textit{Counter terms} $N_{i}(\lambda)$ are expected to be
\textit{factors of automorphy} associated to the fractional linear
transformations of (approximate) \textit{conformal} symmetry $SO(2,4)$($\simeq
SU(2,2)$) associated with (approximate) scale invariance. Along this line, the
Callan-Symanzik type equation for $N_{i}(\lambda)$ involving running coupling
constants and anomalous dimensions should be established.

\item \textit{In the opposite direction} to the conventional renormalization
scheme based on \textit{perturbative expansion method} starting from a
\textit{\textquotedblleft Lagrangian\textquotedblright} (along such a flow
chart as \textquotedblleft Lagrangian\textquotedblright\ $\rightarrow$
perturbative expansion $\rightarrow$ renormalization + OPE), the perturbation
expansion itself should be derived and justified as a kind of asymptotic
analysis within the non-perturbative formulation of renormalization based on
OPE: namely, we advocate such a flow chart as starting from OPE $\rightarrow$
renormalization $\rightarrow$ perturbative method as asymptotic expansion
$\rightarrow$ \textquotedblleft Lagrangian\textquotedblright\ determined by
$\Gamma_{1PI}$ \& renormalizability (= finite generation property).

\item More detailed mathematical connections should be clarified among
\textit{nuclearity condition, renormalizability, renormalization conditions,
renormalization group to shift renormalization point} and \textit{broken scale
invariance inherent to local subalgebras} $\mathcal{A}(\mathcal{O})$
\textit{of type III} from the viewpoint of \textit{non-standard analysis}.
\end{enumerate}

\section*{Acknowledgement}

I would like to express my sincere thanks to Prof. M. Ohya for the invitation
to the pioneering and inspiring International Conference, QBIC2008, and to
Prof. L. Accardi and to Prof. T. Hida for their encouragements.

\end{document}